\documentclass[12pt,preprint]{aastex}
\usepackage{graphicx}
\begin{document}
\title{Nonsingular density profiles of dark matter halos and
Strong gravitational lensing}
\author{Da-Ming Chen}
\affil{National Astronomical Observatories, Chinese Academy of
Sciences, Beijing 100012, China}
\begin{abstract}
We use the statistics of strong gravitational lenses to
investigate whether mass profiles with a flat density core are
supported. The probability for lensing by halos modeled by a
nonsingular truncated isothermal sphere (NTIS) with image
separations greater than a certain value (ranging from $0''$ to $10''$) 
is calculated. NTIS is an analytical model for the
postcollapse equilibrium structure of virialized objects derived
by Shapiro, Iliev, \& Raga. This profile has a soft core and
matches quite well with the mass profiles of dark matter-dominated
dwarf galaxies deduced from their observed rotation curves. It
also agrees well with the NFW (Navarro-Frenk-White)
profile at all radii outside of a few NTIS core radii.
Unfortunately, comparing the results with those for singular
lensing halos (NFW and SIS+NFW) and strong lensing observations,
the probabilities for lensing by NTIS halos are far too low. As
this result is valid for any other nonsingular density profiles
(with a large core radius), we conclude that nonsingular density
profiles (with a large core radius) for CDM halos are ruled out by
statistics of strong gravitational lenses.
\end{abstract}

\section{Introduction}

The concordant, currently dark energy-dominated, spatially flat
cold dark matter (CDM) cosmology is so successful that we are now
said to be in an era of ``precision cosmology"
\citep{peebles02,ostriker04}. In this concordant cosmology, dark
energy (DE), mainly introduced to close the universe and to
explain the accelerating expansion of the universe, attracts much
attention in the realms of astrophysics and theoretical
physics. The simplest candidate for DE is the cosmological constant,
and its more general form, known as quintessence, is a cosmic
scalar field minimally coupled with the usual matter
\citep{caldwell,ma,peebles03}. The generalized Chaplygin gas, as a
unification of dark matter and dark energy, was recently proposed
and can be constrained by observations \citep{pad,zhu04}. In other
cosmological models, DE is replaced by certain possible
mechanisms, such as brane world cosmologies \citep{rsa,rsb} and
the Cardassian expansion model
\citep{freese02,zhu02,zhu03,zhu04a}. Despite its success, the
standard CDM theory of cosmic structure formation has several
problems, which exist mostly in the small-scale regime. For
example, the observed rotation curves of dark-matter-dominated
dwarf and low surface brightness (LSB) disk galaxies tend to favor
mass profiles with a flat density core
\citep[e.g.,][]{salucci00,gentile04}, unlike the singular profiles
of the CDM $N$-body simulations (e.g., Navarro, Frenk \& White 1997,
NFW profiles; Moore et al. 1999; Jing 2000; Jing \& Suto 2002) and 
that favored by baryon cooling models (i.e., singular isothermal
sphere, SIS profile). While there are debates on whether the 
observed data were resolved well enough to indicate a soft core
\citep{van,march}, quite recent $N$-body simulations of CDM with
higher and higher force and mass resolution still favor cuspy halo
profiles \citep{diemand,fukushige,navarro04,tasitsiomi,wambs04}.

Recently, an analytical model was presented for the post-collapse
equilibrium structure of virialized objects that condense out of a
low-density cosmological background universe, with or without
a cosmological constant \citep{shapiro99,Iliev01}. The model is
based on the assumption that cosmological halos form from the
collapse and virialization of `top-hat' density perturbations, and
are spherical, isotropic and isothermal. According to the authors,
this predicts a unique, non-singular, truncated
isothermal sphere (NTIS) and provides a simple physical clue
about the existence of soft cores in halos of cosmological origin.
This NTIS model is claimed to be in good agreement with observations of
the internal structure of dark matter dominated halos on scales
ranging from dwarf galaxies to X-ray clusters. In particular, it
matches quite well the mass profiles of dark matter dominated
dwarf galaxies deduced from their observed rotation curves
\citep{shapiro04}. Quite recently, the NTIS model was revisited by
using the self-interacting dark matter (SIDM) hypothesis
\citep{kyungjin}. There are other efforts for analytical
derivations of the density profile; e.g., \citet{hansen04} derived
the bound on the central density slope of -1 analytically (as
found numerically by NFW). This is done by a simple solution to
the Jeans equation, which is valid under the assumption that both
the central density profile and the phase-space-like density are
exact power laws. However, this work did not clearly give a density
core.

To investigate whether there is a soft core at the center of each
CDM halo, we use another independent and robust tool,
gravitational lensing, which has been widely used to detect the
mass distributions in the Universe
\citep{turner,schne,wambs95,wu,barte96,wu00,barte01,chen01,keeton2001a,
keeton2001b,xue01,keeton2002,keeton2003,pen03,schne03,
keeton2004,wu04,zhang04b} and the dark energy density and its
equation of state
\citep{fukugita90,fukugita91,turner90,krauss92,maoz93,
kochanek95,kochanek96,falco98,cooray99,waga99,sarbu01,dev04}.
Recently, motivated by the current largest homogeneous sample of
lensed quasars, coming from the radio Cosmic Lens All-Sky
Survey\citep[CLASS; ][]{myers,browne03}, an extension of the
earlier Jodrell Bank/Very Large Array Astrometric Survey
\citep[JVAS; ][]{patnaik92,king99}, much work has been devoted to
statistics of strong lensing to constrain the cosmological
parameters and mass distribution
\citep{chae02,li02,oguri02,chae03,li03,oguri03a,oguri03b,oguri04,
wj04,mitchell05,sereno05,zhang05}. Statistics of strong
gravitational lensing by halos with a soft core was studied quite
early \citep{hinshaw,chiba,cheng}, but no observational sample
suitable for analysis could be used in their work; in particular,
they did not use their cored density profile to fit the observed
rotation curves of the dwarf and LSB disk galaxies. In strong
gravitational lensing of quasars, the separation between multiple
images is the most important observable.  In some instances,
asymmetry and shear of lens halos affect the properties of images
considerably, but statistically, image separations are mainly
determined by the potential wells of the host halos characterized
by the slopes of their density profiles. Thus, the probabilities for
lensing by NTIS halos with image separations greater than
$\Delta\theta$ provide us with an independent and powerful probe
of the existence of soft cores of CDM halos. Our calculations are
performed in a concordant, spatially flat $\Lambda$CDM cosmology
favored by the \textit{Wilkinson Microwave Anisotropy Probe} 
\citep[\textit{WMAP};][]{bennett03} plus large-scale structure (LSS) data 
from the Sloan Digital Sky Survey \citep[SDSS;][]{tegmark04a,tegmark04b}, 
with matter density parameter $\Omega_\mathrm{m}=0.3$, galaxy
fluctuation amplitude $\sigma_8=0.9$, and Hubble parameter
$h=0.72$. For comparison, we also give the results when NTIS is
replaced by SIS+NFW and NFW profiles.

\section{Lensing Probabilities}
In what follows, we use a reduced mass of a halo defined as
$M_{15}=M/(10^{15}\mathrm{h}^{-1}M_{\sun})$. The NTIS model is
based on the assumption that cosmological halos form from the
collapse and virialization of top-hat density perturbations and
are spherical, isotropic, and isothermal
\citep{shapiro99,Iliev01}.  The well-fitted density profile is
given by \citep{shapiro99,Iliev01,shapiro04}
\begin{equation}
\rho(r)=\rho_0\left(\frac{A}{a^2+r^2/r_0^2}
-\frac{B}{b^2+r^2/r_0^2}\right),
\end{equation}
where $A=21.38$, $B=19.81$, $a=3.01$, and $b=3.28$. The central value
of the density profile is $\rho_0=1.8\times
10^4\rho_\mathrm{c}(z_\mathrm{coll})$, where
$\rho_\mathrm{c}(z_\mathrm{coll})$ is the critical density of the
Universe at the epoch of halo collapse with redshift
$z_\mathrm{coll}$. The small core radius depends on mass $M$ and
$z_\mathrm{coll}$ and is given by
\begin{equation}
r_0=0.115(M/\rho_0)^{1/3}=6.73\times
10^{-2}M_{15}^{1/3}[\Omega_\mathrm{m}(1+z_\mathrm{coll})^3
+\Omega_{\Lambda}]^{-1/3}h^{-1}\mathrm{Mpc}, \label{r0}
\end{equation}
where we have used
$\rho_\mathrm{c}(z)=\rho_\mathrm{c}(0)[\Omega_\mathrm{m} (1+z)^3
+\Omega_{\Lambda}]$ (with $\Omega_\mathrm{m}=0.3$ and
$\Omega_\Lambda=0.7$ in later calculations) and
$\rho_\mathrm{c}(0)=1.88\times
10^{-29}h^2\mathrm{g}\mathrm{cm}^{-3}=2.777\times
10^{11}h^2M_\sun\mathrm{Mpc}^{-3}$. It should be pointed out that
the original formula for $r_0$, equation (88) in
\citet{shapiro04}, is $r_0=1.51\times 10^{-3}(M/\rho_0)^{1/3}$ (we
write $M_{200}$ as $M$, which is discussed in section 3).
This is a clerical error, but we have checked that their
subsequent results are not affected by this wrong formula.

The gravitational lens equation is
$\eta=D_\mathrm{S}\xi/D_\mathrm{L}-D_\mathrm{LS}\hat{\alpha}$,
where $\eta$ and $\xi$ are the physical positions of a source in
the source plane and an image in the image plane, respectively,
$\hat{\alpha}$ is the deflection angle, and $D_\mathrm{L}$,
$D_\mathrm{S}$, and $D_\mathrm{LS}$ are the angular diameter
distances from observer to lens, observer to source, and lens to
source, respectively. By defining dimensionless positions
$y=D_\mathrm{L}\eta/D_\mathrm{S}r_0$ and $x=\xi/r_0$, and
dimensionless angle
$\alpha=D_\mathrm{L}D_\mathrm{LS}\hat{\alpha}/D_\mathrm{S}r_0$,
the lens equation is then \citep{shapiro04}
\begin{equation}
y=x-\frac{2ab\kappa_c}{(Ab-Ba)}\left(A\sqrt{a^2+x^2}
-B\sqrt{b^2+x^2}-Aa+Bb\right), \label{lenseq1}
\end{equation}
where $\kappa_c$ is the central convergence:
\begin{equation}
\kappa_c=\frac{\Sigma(\xi=0)}{\Sigma_\mathrm{crit}}=\frac{\pi\rho_0
r_0}{\Sigma_\mathrm{crit}}\left(\frac{A}{a}-\frac{B}{b}\right),
\label{kc1}
\end{equation}
where $\Sigma_\mathrm{crit}=c^2D_\mathrm{S}/4\pi
GD_\mathrm{L}D_\mathrm{LS}$ is the critical surface density.

It is well known that generally, for any spherically symmetric
density profiles of lensing halos, multiple images can be produced
only if the central convergence is greater than unity
\citep{schne}. When $\kappa_c\leq 1$, only one image is produced.
Note that even if $\kappa_c>1$ is satisfied, multiple images can
occur only when the source is located within
$y_\mathrm{cr}=y(x_\mathrm{cr})$, where $x_\mathrm{cr}$ is
determined from the lensing equation (eq. [\ref{lenseq1}]) with 
$dy/dx=0$
for $x<0$ (this is similar to lensing by NFW halos). For a singular
density profile such as the SIS and NFW profiles, the central value 
is divergent,
so $\kappa>1$ is always satisfied, and multiple images can be
produced for any given mass. For density profiles with a finite
soft core such as the NTIS profile, however, the condition 
$\kappa>1$ requires
that only halos with mass greater than a certain value (determined
by $\kappa_c=1$) can produce multiple images. This is clearly
shown in Figure 1, where three curves for $\kappa_c=1.1, 1.05$, and
$1.0$ are plotted, and when $\kappa_\mathrm{c}=1.0$, only one
image is produced. In lensing statistics, this requirement will
limit the populations of lensing halos to quite a small fraction.
Such a conclusion is valid for any lensing halos with a finite
soft core, which is discussed in detail later.

When quasars at redshift $z_{\mathrm{s}}$ are lensed by foreground
CDM halos of galaxies and clusters of galaxies, the lensing
probability for image separations larger than $\Delta\theta$ is
\citep{turner,schne}
\begin{equation}
P(>\Delta\theta)= \int\mathcal{P}(z_s)dz_s\int^{z_{\mathrm{s}}}_0
\frac{dD_{\mathrm{L}}^\mathrm{p}(z)}
{dz}dz\int^{\infty}_0\bar{n}(M,z)\sigma(M,z)B(M,z)dM,
\label{prob1}
\end{equation}
where $\mathcal{P}(z_\mathrm{s})$ is the redshift distribution for
quasars approximated by a Gaussian model with a mean of 1.27 and
a dispersion of 0.95 \citep{helbig1999,marlow2000,chae02,myers},
$D_{\mathrm{L}}^\mathrm{p}(z)$ is the proper distance from the
observer to the lens located at redshift $z$, $\bar{n}(M,z)$ is
the physical number density  of virialized dark halos of masses
between $M$ and $M+dM$, and  $B(M,z)$ is the magnification bias.
The physical number density $\bar{n}(M,z)$ is related to the 
comoving number density $n(M,z)$
by $\bar{n}(M,z)=n(M,z)(1+z)^3$; the latter is originally given by
\citet{press74}, and the extended version is
$n(M,z)dM=(\rho_0/M)f(M,z)dM$, where $\rho_0$ is the current mean
mass density of the universe and
\begin{equation}
f(M,z)=(0.315/M)(d\ln\Delta_{\mathrm{z}}/d\ln
M)\exp(-|\ln(\Delta_{\mathrm{z}}/1.68)+0.61|^{3.8})
\end{equation}
is the mass
function for which we use the expression given by \citet{jenki}.
In this expression, $\Delta_{\mathrm{z}}=\delta_c(z)/\Delta(M)$,
in which $\delta_c(z)$ is the overdensity threshold for spherical
collapse at redshift $z$ and $\Delta(M)$ is the rms of the present
variance of the fluctuations in a sphere containing a mean mass
$M$. The overdensity threshold is given by $\delta_c(z)=1.68/D(z)$
for the $\Lambda$CDM cosmology \citep{nfw97}, where
$D(z)=g[\Omega(z)]/[g(\Omega_{\mathrm{m}})(1+z)]$ is the linear
growth function of the density perturbation \citep{carroll}, in
which $g(x)=0.5x(1/70+209x/140-x^2/140+x^{4/7})^{-1}$ and
$\Omega(z)=\Omega_{\mathrm{m}}(1+z)^3
/[1-\Omega_{\mathrm{m}}+\Omega_{\mathrm{m}}(1+z)^3]$. When we
calculate the variance of the fluctuations $\Delta^2(M)$, we use
the fitting formulae for the CDM power spectrum $P(k)=AkT^2(k)$ given
by \citet{eisen}, where $A$ is the amplitude normalized to
$\sigma_8=\Delta(r_{\mathrm{M}}=8h^{-1}\mathrm{Mpc})$, given by
observations. The cross section for lensing is $\sigma(M,z)=\pi
y_\mathrm{cr}^2r_0^2\vartheta(M-M_{\mathrm{min}})$, where
$\vartheta(x)$ is a step function, and $M_{\mathrm{min}}$ is the
minimum mass of halos above which lenses can produce images with
separations greater than $\Delta\theta$. 
The minimum mass $M_{\mathrm{min}}$ can be
derived directly from the relationship between the mass of a lens
halo and the corresponding image separation, as follows. It is
obvious from Figure 1 that the separation between the outer two
images is almost independent of the source position $y$. Thus,
similar to the analysis for NFW profiles \citep{li02}, the image 
separation
$\Delta x(y)$ produced by a NTIS halo for a source at $y$ can be
well approximated by $\Delta x(0)=2x_0$, where $x_0$ is the
Einstein radius determined from the lens equation with $y=0$. The
angular image separation is
\begin{equation}
\Delta\theta=\frac{\Delta
xr_0}{D_\mathrm{L}}=\frac{2x_0r_0}{D_\mathrm{L}}.
\end{equation}
Then, from this equation and equation(\ref{r0}) we have
\begin{equation}
M_{\mathrm{min}}=1.26\times 10^{12}\left(\frac{1}{x_0}\right)
\left(\frac{\Delta\theta}
{1''}\right)\left(\frac{D_\mathrm{L}}{c/H_0}\right)^3
[\Omega_\mathrm{m}(1+z)^3+\Omega_\Lambda] \mathrm{h}^{-1}M_{\sun},
\end{equation}
where  $c/H_0=2997.9h^{-1}$Mpc is the present Hubble radius. The
magnification bias $B(M,z)$ is calculated numerically with
\begin{equation}
B(M,z)=(2/y_\mathrm{cr})\int^{y_\mathrm{cr}}_0dyy[\mu(y)]^{1.1}
\end{equation}
given by \citet{oguri02}, where $\mu(y)$ is the total
magnification of the two outer images for a source at $y$; this
can also be computed numerically.

The numerical results of Eq. (\ref{prob1}) for NTIS lens halos are
plotted in figure 2 (\textit{thin solid line}). For comparison, 
the survey
results of JVAS/CLASS and the predicted probability for lensing by
SIS + NFW and NFW profiles are also shown. A subset of 8958 sources from
the combined JVAS/CLASS survey  form a well-defined statistical
sample containing 13 multiply imaged sources (lens systems)
suitable for analysis of the lens statistics
\citep{myers,browne03,patnaik92,king99}.  The observed lensing
probabilities can be easily calculated \citep{chenb,chenc} by
$P_{\mathrm{obs}}(>\Delta\theta)=N(>\Delta\theta)/8958$, where
$N(>\Delta\theta)$ is the number of lenses with separation greater
than $\Delta\theta$ in 13 lenses.
The observational probability $P_{\mathrm{obs}}(>\Delta\theta)$ is 
plotted as a histogram in
Figure 2. In the two-population model SIS + NFW, the galaxy-size
and the cluster-size lens halos are approximated by SIS and NFW 
profiles, respectively \citep{sarbu01,li02,
chena,chenb,chenc,chend,zhang2004}. In the one-population model
NFW, lens halos with different sizes are all approximated by NFW
profiles
\citep{li02}; this is similar to the NTIS model. The theoretically
predicted lensing probabilities shown in Figure 2 are calculated
separately for three cases (NTIS, SIS+NFW and NFW) according to
equation (\ref{prob1}) . The differences in lensing probability
distributions for the three cases arise from their different values
for the lensing cross section $\sigma(M,z)$ and magnification bias
$B(M,z)$, since these two quantities are determined uniquely from
the corresponding density profile . Here we have recalculated the
lensing probabilities for SIS + NFW and NFW profiles according to equation
(\ref{prob1}). Since the density profiles, lensing equations, and
lensing cross sections for SIS and NFW profiles have been discussed many
times in the literature, here we only give the final results of
lensing probabilities for these two models \citep[for details about SIS+NFW and
NFW profiles, see][and
references therein]{li02,chenc}.

\section{Discussion and Conclusions}\label{dis}

One can see clearly from Figure 2 that probability for lensing by
NTIS halos with an image separation greater than $\Delta\theta$ is
far too low to match the observational results of CLASS/JVAS; it
is even much less  than that for NFW lenses. We thus conclude
that, at least, NTIS as a model to approximate density profiles of
dark matter halos is ruled out by statistical strong lensing.
Within the framework of statistics of strong lensing as displayed
in the literature, no mechanism can save this model.

In fact the above conclusion is general for any mass profile with
a flat soft core characterized by the core density
$\rho_\mathrm{core}$ and core radius $r_\mathrm{core}$ (these two
parameters are determined by rotation curves of dark matter
dominated dwarf and LSB disk galaxies). To see this, we revisit
another such density profile, an isothermal sphere with a soft
core (cored isothermal sphere, CIS):
$\rho(r)=\sigma_\mathrm{v}^2/2\pi G(r^2+r_\mathrm{core}^2)$
\citep{hinshaw,chiba,cheng}, where $\sigma_\mathrm{v}$ is the
one-dimensional velocity dispersion. The lens equation is
\begin{equation}
y=x-2\kappa_\mathrm{c}^{\mathrm{CIS}}(\sqrt{x^2+1}-1)/x,
\label{lenseq2}
\end{equation}
 where $y$ and $x$ are
defined in the same way as in equation (\ref{lenseq1}) (here
$x=\xi/r_\mathrm{core}$) and
\begin{equation}
\kappa_\mathrm{c}^{\mathrm{CIS}}=\sigma_\mathrm{v}^2/2Gr_\mathrm{core}
\Sigma_\mathrm{crit} \label{kc2}
\end{equation}
is the central convergence. Similarly, multiple images can be
produced if and only if $\kappa_\mathrm{c}^{\mathrm{CIS}}>1$. The
corresponding lens equation is plotted in Figure 3 for three
different values of $\kappa_\mathrm{c}^{\mathrm{CIS}}$; the curves
are quite similar to the NTIS model, even though their density
profiles seem quite different.

It is not difficult to understand the extremely low value of the
probability for lensing by dark halos with a flat soft core. In
fact, in our previous calculations, for the NTIS profile,
$\kappa_\mathrm{c}$ can be written in the form
\begin{equation}
\kappa_\mathrm{c}=475.7\frac{\rho_\mathrm{c}^{2/3}(z_\mathrm{coll})}
{\Sigma_\mathrm{crit}}M_{200}^{1/3}=3.65M^{1/3}_{15}
[\Omega_\mathrm{m}(1+z)^3+\Omega_{\Lambda}]^{2/3}
\frac{D_\mathrm{R}}{c/H_0}, \label{kc3}
\end{equation}
where $D_\mathrm{R}=D_\mathrm{L}D_\mathrm{LS}/D_\mathrm{s}$ . We
have defined, as usual, the mass of a dark matter halo to be
$M=M_{200}=4\pi\int_0^{r_{200}}\rho(r)r^2dr$ ($r_{200}$ has its
usual definition) and assume that the redshift $z_\mathrm{coll}$
by which the dark halo collapsed  is equal to the lens location
redshift $z_\mathrm{L}$ and that such an assumption will not affect our
results \citep{shapiro04}. For a source at $z_\mathrm{s}=3.0$ and
the lens at $z_\mathrm{L}=0.5$, equation (\ref{kc3}) gives
$\kappa_\mathrm{c}=1.08M_{15}^{1/3}$. In this case,
$\kappa_\mathrm{c}=1$ implies $M\sim M_{15}$, and this means that
for this typical lens system, multiple images can be produced only
when the lens mass is higher than
$10^{15}\mathrm{h}^{-1}M_{\sun}$, which is the typical size of
galaxy clusters. By setting $\kappa_\mathrm{c}=1$ in equation
(\ref{kc3}), the minimum lens mass for producing multiple images
is determined as a function of both $z_\mathrm{L}$ and
$z_\mathrm{s}$. In Figure 4, we plot the mass
$M(\kappa_\mathrm{c}=1)$ (measured in $M_{15}$) as a function of
$z_\mathrm{s}$ for three given values of $z_\mathrm{L}$: 0.1, 0.5,
and 1.0. For a given $z_\mathrm{L}$, the minimum
mass decreases with increasing $z_\mathrm{s}$. Since the mean
value of the redshift of the sources is $\langle
z_\mathrm{s}\rangle=1.27$ in our calculations, the actual mean
value of the mass for producing multiple images (for lenses at
$z_\mathrm{L}=0.5$) is larger than
$10^{15}\mathrm{h}^{-1}M_{\sun}$. It is also obvious in Figure 4
that in most cases, multiple images can be produced only when
the halo mass is larger than $10^{15}\mathrm{h}^{-1}M_{\sun}$. From
equation (\ref{prob1}), we know that the lensing probability is
determined by the number density of dark halos and the cross section.
As pointed out by \citet{shapiro04}, an NFW profile would produce
more multiple-image lenses than an NTIS one at relatively lower masses,
and this trend is reversed at higher masses. Statistically,
however, it is the requirement of $\kappa_\mathrm{c}>1$ and thus
the existence of a large core radius that strongly limits the
populations of lensing halos [number density $\bar{n}(M,z)$] to
quite a small fraction. Thus, the extremely low value of the
probability for lensing by NTIS halos arises from the quite low
value of the corresponding number density of galaxy clusters.
Furthermore, the lensing cross section $\sigma\sim
y_\mathrm{cr}^2$, and from Figure 1 and Figure 3 we see that
$y_\mathrm{cr}$ is quite sensitive to
$\Delta\kappa_\mathrm{c}=\kappa_\mathrm{c}(>1)-1$ rather than
$\kappa_\mathrm{c}$ itself. Therefore, a slight change in
$\kappa_\mathrm{c}\sim M^{1/3}$ will result in a large change both
in $y_\mathrm{cr}$ and $x_0$ (the Einstein radius), and since
$x_0\sim \Delta\theta$, this explains the quite flat curve for
the NTIS lensing probability in Figure 2 (\textit{thin solid line}). 
Namely, the
insensitivity of NTIS lensing probability to $\Delta\theta$
reflects the fact that within the image separations of a few
arcseconds, the lens mass and thus the corresponding number
density of dark halos around that mass change only a little.
Similarly, for a CIS profile, after solving the equations
\begin{eqnarray}
M&=&800\pi\rho_\mathrm{c}r_{200}^3/3 \label{mball} \\
&=&4\pi\int^{r_{200}}_0drr^2 \sigma_\mathrm{v}^2/2\pi
G(r^2+r_\mathrm{core}^2)\\
&\approx&2\sigma_\mathrm{v}^2(r_{200}-\pi r_\mathrm{core}/2)/G
\label{sigmav}
\end{eqnarray}
for $r_{200}$ and $\sigma_\mathrm{v}^2$,
$\kappa_\mathrm{c}^{\mathrm{CIS}}$ can be related to dark halo
mass $M$. The above analysis for the NTIS profile then is also true
for the CIS model, if core radius $r_\mathrm{core}$ is large enough to
be able to fit the observed rotation curves of the dwarf and LSB
disk galaxies.

Note that our conclusion that a CIS model as a cored mass profile is
ruled out by statistical strong gravitational lensing seems
inconsistent with previous works \citep{hinshaw,chiba,cheng},
since these authors used this model to constrain cosmological
parameters. A simple analysis, however, shows us that there is no
discrepancy. The central convergence can be expressed in terms of
the mass and the core radius of a lens halo. For NTIS, from
equation (\ref{r0}) and equation (\ref{kc1}), we have
$\kappa_\mathrm{c}\propto M/r_0^2$; for CIS, from equation
(\ref{kc2}), equation (\ref{mball}) and equation (\ref{sigmav}) we
have (approximately) $\kappa_\mathrm{c}^{\mathrm{CIS}}\propto
M^{2/3}/r_\mathrm{core}$. A larger $r_\mathrm{core}$ needs a larger
$M$ to ensure $\kappa_\mathrm{c}^{\mathrm{CIS}}\geq 1$, or, for a
fixed value of mass $M$, an appropriate smaller value of
$r_\mathrm{core}$ would ensure
$\kappa_\mathrm{c}^{\mathrm{CIS}}\geq 1$. Therefore, if
$r_\mathrm{core}$ is not determined by currently observed rotation
curves of the dwarf and LSB disk galaxies, but rather is
adjustable \citep{hinshaw,chiba,cheng}, then no discrepancy
appears. A more realistic mass profile should not be divergent at
the center (cusp), that is, it should have a flat core, but the
core radius $r_\mathrm{core}$ should be small enough to ensure that 
the lensing probability matches the observations of CLASS/JVAS.

\begin{acknowledgements}
I thank the anonymous referee for quite useful suggestions. This
work was supported by the National Natural Science Foundation of
China under grant 10233040.
\end{acknowledgements}

\clearpage
\begin{figure}
\epsscale{0.60} \plotone{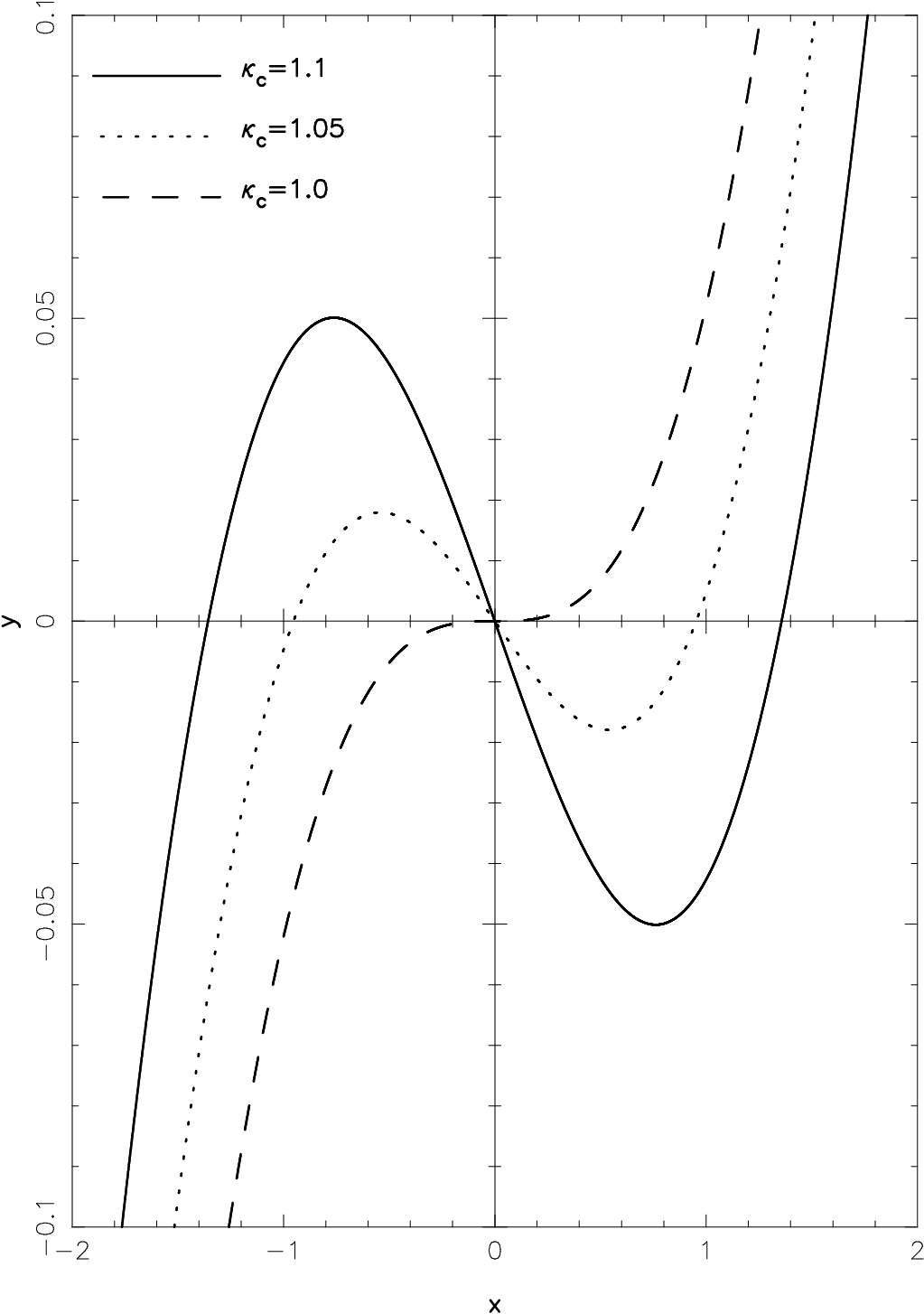} \caption{Lens equation for NTIS
halos, plotted according to equation (\ref{lenseq1}). Coordinates
$y$ and $x$
are dimensionless positions of the source and the image,
respectively. Curves for three values of $\kappa_\mathrm{c}=$1.1
(\textit{solid curve}), 1.05 (\textit{dotted curve}) and 1.0 
(\textit{dashed curve}) are plotted.
When $\kappa_\mathrm{c}=1$, only one image is created.}
\end{figure}
\clearpage

\clearpage
\begin{figure}
\epsscale{0.80} \plotone{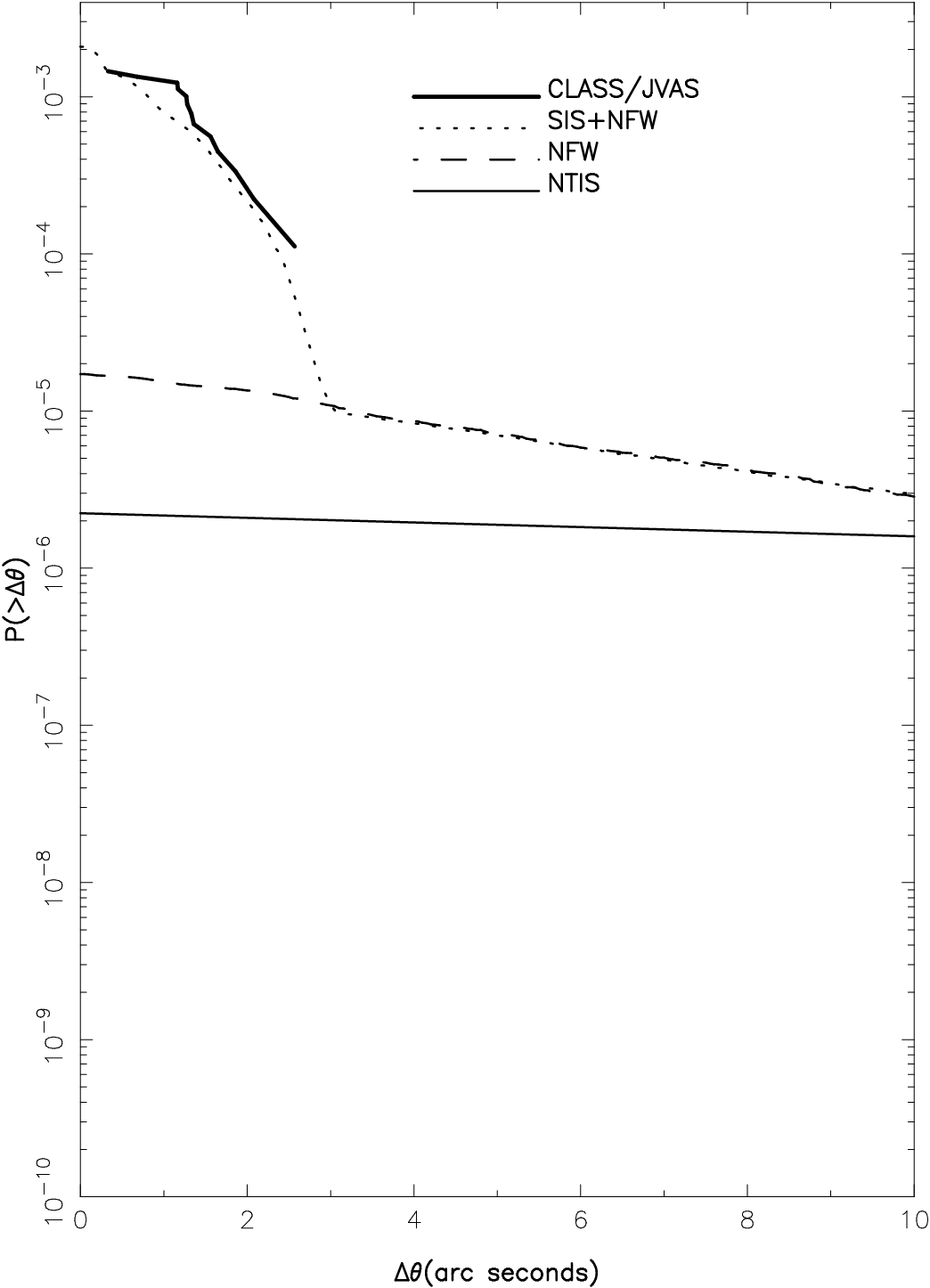} \caption{Predicted probability
for an image separation angle greater than $\Delta\theta$ for
lensing by NTIS halos (\textit{thin solid line}). For comparison, the survey
results of JVAS/CLASS (\textit{thick histogram}), the predicted
probability for lensing by SIS+NFW (\textit{dotted line}) and NFW 
(\textit{dashed line}) profiles are also shown.}
\end{figure}
\clearpage

\clearpage
\begin{figure}
\epsscale{0.80} \plotone{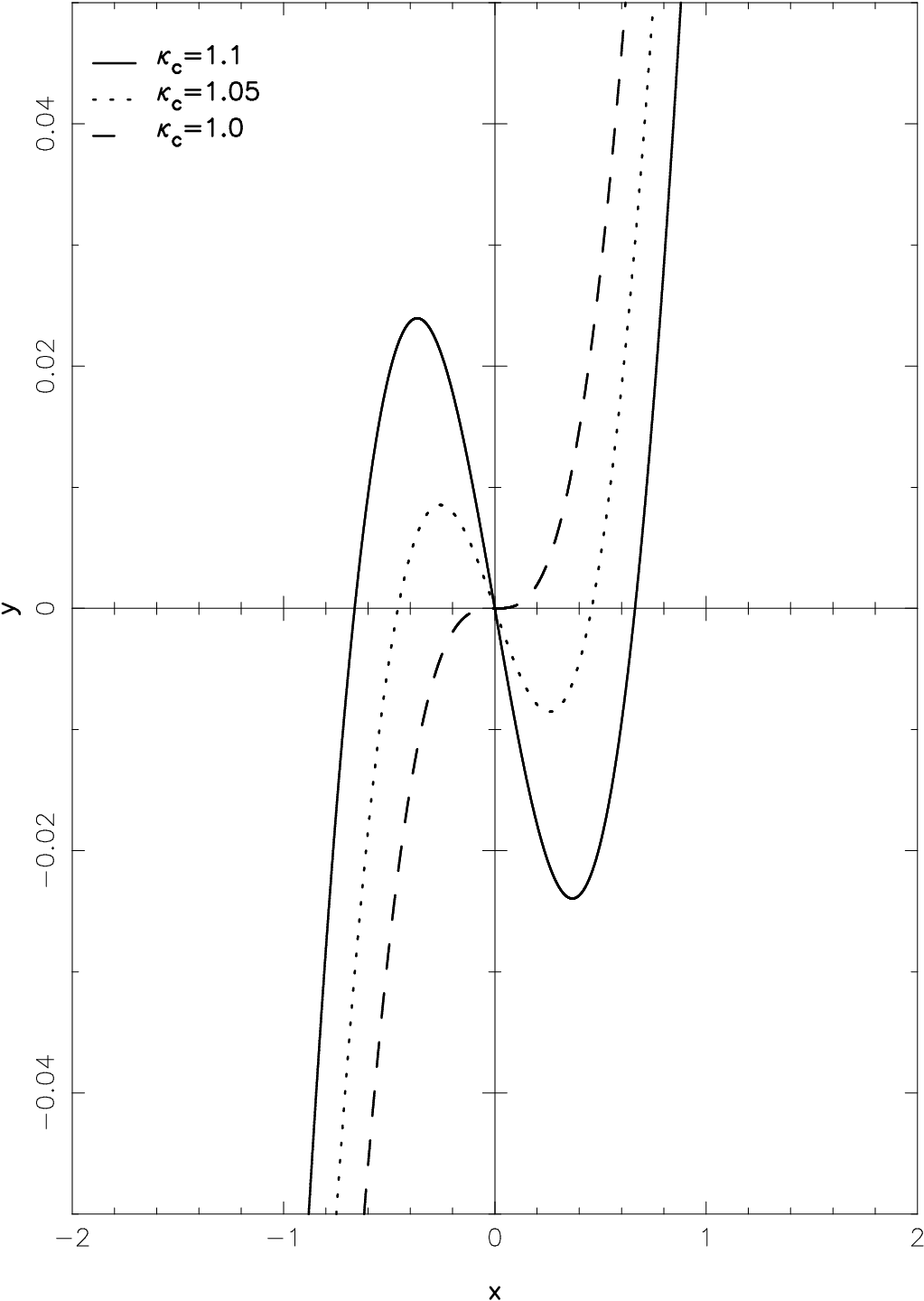} \caption{Lens equation for CIS
halos, plotted according to equation (\ref{lenseq2}). Coordinates
$y$ and $x$
are dimensionless positions of the source and the image,
respectively. Curves for three values of $\kappa_\mathrm{c}=$1.1
(\textit{solid curve}), 1.05 (\textit{dotted curve}) and 1.0 
(\textit{dashed curve}) are plotted.
When $\kappa_\mathrm{c}=1$, only one image is created.}
\end{figure}
\clearpage

\clearpage
\begin{figure}
\epsscale{0.8} \plotone{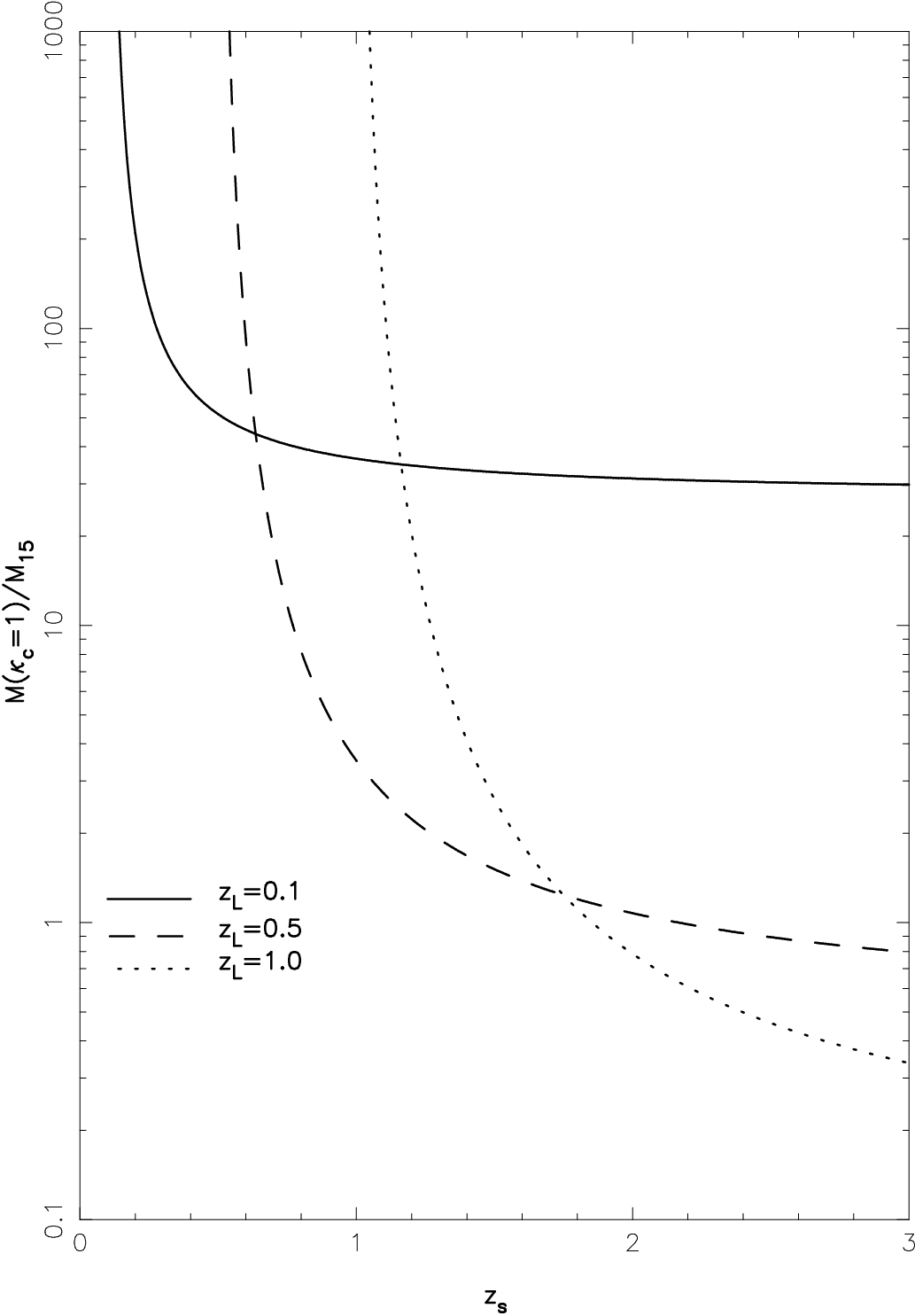} \caption{Lens mass
$M(\kappa_\mathrm{c}=1)$ as a function of $z_\mathrm{s}$,
determined by setting $\kappa_\mathrm{c}=1$ in equation
(\ref{kc3}) for given values of $z_\mathrm{L}$. Curves for three
typical values of $z_\mathrm{L}=$0.1 (\textit{solid curve}), 0.5 (\textit{dashed curve}) and 1.0
(\textit{dotted curve}) are plotted.}
\end{figure}

\end{document}